\documentclass[superscriptaddress]{revtex4-2}

\usepackage{amsmath}
\usepackage{mathtools}
\usepackage{mathpazo} 
\usepackage{graphicx}

\usepackage{color}
\usepackage{graphicx} 
\usepackage{float}
\usepackage{epstopdf}%
\usepackage{placeins}
\usepackage{dsfont}
\usepackage{xcolor}

\usepackage{dblfloatfix}

\usepackage{subfloat}

\newcommand {\be} {\begin{equation}} 
\newcommand {\ba}{\begin{eqnarray}} 
\newcommand {\ee} {\end{equation}} 
\newcommand{\ea} {\end{eqnarray}}
\newcommand\cev[1]{\overleftarrow{#1}}
\newcommand\vvec[1]{\overrightarrow{#1}}

\begin{document}


\title{Quantifying information via Shannon entropy in spatially structured optical beams}



\author{Maria Solyanik-Gorgone}
\email[]{msolyanik@gwu.edu}
\affiliation{Department of Electrical and Computer Engineering, The George Washington University, Washington, DC 20052, USA}

\author{Jiachi Ye}
\affiliation{Department of Electrical and Computer Engineering, The George Washington University, Washington, DC 20052, USA}

\author{Mario Miscuglio}
\email[]{mmiscuglio@gwu.edu}
\affiliation{Department of Electrical and Computer Engineering, The George Washington University, Washington, DC 20052, USA}

\author{Andrei Afanasev}
\email[]{afanas@gwu.edu}
\affiliation{Department of Physics, The George Washington University, Washington, DC 20052, USA}

\author{Alan Willner}
\email[]{willner@usc.edu}
\affiliation{Department of Electrical Engineering at University of Southern California, Los Angeles, California 90089, USA}

\author{Volker J. Sorger}
\email[]{sorger@gwu.edu}
\affiliation{Department of Electrical and Computer Engineering, The George Washington University, Washington, DC 20052, USA}


\date{\today}

\begin{abstract}
While information is ubiquitously generated, shared, and analyzed in a modern-day life, there is still some controversy around the ways to asses the amount and quality of information inside a noisy optical channel. A number of theoretical approaches based on, e.g., conditional Shannon entropy and Fisher information have been developed, along with some experimental validations. Some of these approaches are limited to a certain alphabet, while others tend to fall short when considering optical beams with non-trivial structure, such as Hermite-Gauss, Laguerre-Gauss and other modes with non-trivial structure. Here, we propose a new definition of classical Shannon information via the Wigner distribution function, while respecting the Heisenberg inequality. Following this definition, we calculate the amount of information in a Gaussian, Hermite-Gaussian, and Laguerre-Gaussian laser modes in juxtaposition and experimentally validate it by reconstruction of the Wigner distribution function from the intensity distribution of structured laser beams. We experimentally demonstrate the technique that allows to infer field structure of the laser beams in singular optics to assess the amount of contained information. Given the generality, this approach of defining information via analyzing the beam complexity is applicable to laser modes of any topology that can be described by 'well-behaved' functions. Classical Shannon information defined in this way is detached from a particular alphabet, i.e. communication scheme, and scales with the structural complexity of the system. Such a synergy between the Wigner distribution function encompassing the information in both real and reciprocal space, and information being a measure of disorder, can contribute into future coherent detection algorithms and remote sensing.
\end{abstract}


\maketitle

\section{Introduction}   \label{sec:intro}
An electromagnetic field is a fundamental physical carrier of information. It is capable of reliably transmitting a modulated signal and collecting information about the propagation channel itself. With relevance to this IT-driven age, the two longstanding goals in information processing are; (i) achieving higher channel capacity (i.e. throughput), and (ii) (pre)processing of collected information. However, the fundamental challenge of rigorous qualification and quantification of information in EM waves still remains a topic of debate including both physical and even semi-philosophical notions. Information theory stands out from most of other approaches in physics. Being a higher level of abstraction, it focuses on a configuration of a system under consideration in the context of its prehistory, similarly to Thermodynamics, without attachment to a particular class of objects under study in its axiomatics. Out of the broad scope of studies where information theory has a potential to contribute, at this point we exemplary explore its applications to Electrical Engineering, and Signal processing.

One way of defining information is associating it with the presence of \emph{distinctive features}. For instance, human speech can carry up to $2^{14}$ distinctive sounds (due to 14 binary distinctive features, e.g. \cite{jakobsonm}), only a small subset of which is realized in any particular known language. The amount of information a human can transmit per unit sentence containing a fixed number of words, is indirectly correlated with the amount of distinctive features the language can handle (i.e. language capacity). Translating this into optics, it has been understood that a monochromatic plane-wave photon in free space can carry a rather limited amount of distinctive features (i.e. polarization and wavelength), providing up to one bit of information per photon. Naturally,  if one generates photons with up to one bit of information, one is also able to detect only 1-bit information per unit carrier. To use an analogy; consider "a marine biologist casting a fishing net with two inches wide meshes for exploring the life on the ocean, naturally one should not be surprised finding only sea-creature larger than two inches long" \cite{eddington1939philosophy}. This stumbling block has been shifted with the seminal work by Allen \emph{et al.} \cite{Allen:1992zz} where it was experimentally confirmed that laser beams are capable of carrying a well-defined orbital angular momentum (OAM).  An ability of such laser modes to carry theoretically unbounded amount of information per photon, e.g. \cite{mair2001entanglement}, dramatically expands the EM-field's "language capacity".

There are several approaches to assess a signal's information capacity developed in modern information theory, e.g. \cite{cover1999elements, vedral2006introduction}. In many cases, information is defined with respect to a particular alphabet, giving up the generality offered by statistics in the foundation of information theory. Several groups used conditional information approach to quantify the signal capacity \cite{restuccia2016comparing}. Here, we introduce the concept of expressing information as a measure of structure in a physical system by applying the Shannon information theory to singular optical beams. We discuss how the Wigner distribution function (WDF) can be taken as a corresponding probability distribution function accounting for partial \emph{quantumness} of a shaped photon source. The important synergy between a comprehensive description of physical systems in their phase-space, delivered by the WDF, and the generalized axiomatics of the information theory, have the potential to conduce to a cumulative approach to high-information-density telecommunications and adaptive signal processing techniques. We validate this theoretical framework by experimentally showing how increased structural complexity of wavefront-shaped optical beams, such as Hermite-Gauss (HG) modes and optical vortices \cite{galvez2006gaussian, vaity2015perfect}, can be analyzed using wavefront sensors. 

Due to the wide scope of this topic, we refer the reader to a short and selective literature review, offered in the appendix.

\begin{figure}
\includegraphics[scale=.7]{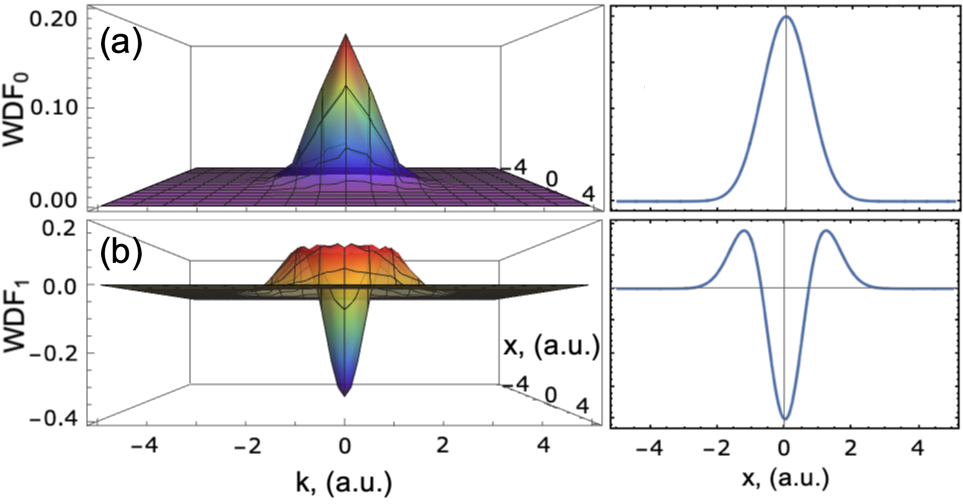}
\vspace{9mm}
\caption{The Wigner distribution functions (WDFs) of one-dimensional (1D) Hermite-Gauss (HG) mode of order zero HG$_0$ (Gaussian) and first order HG$_1$ as functions of $\kappa_x$ - wave vector \eqref{kappa_eq},  and $\bar x$ - coordinate in the beam's transverse plane. One can see that in the case of Gaussian mode the WDF is positive, while for first-order HG-mode there is a negative contribution in the near-zero region of the phase-space.}
\label{fig:experiment}
\end{figure}

\section{The WDF and Classical Information in Optics} \label{sec: WDF}
The WDF belongs to the generalized Cohen's class of dual-domain distributions. It is simultaneously the most complete analytical description of an optical beam, and an observable that can be experimentally measured. It provides access to the spatial beam profile and its Fourier transform. The WDF in one spatial dimension can be defined as
\begin{equation}
W(x,k_x) = \frac{1}{2\pi}\int dy \;u\Big(x+\frac{y}{2}\Big) u^{*} \Big(x-\frac{y}{2}\Big) e^{-ik_x y}
\label{08_13_1}
\end{equation}
where $x$ is the variable in the coordinate space, and $k$ is the corresponding coordinate in reciprocal space. The following integrals have a probabilistic interpretation:
\begin{gather}
|u(k)|^2 = \int dx\; W(x, k);\\
|u(x)|^2 = \int dk\; W(x, k);\\
U_{tot} = \int dx\;dk\; W(x, k);
\end{gather}
where $|u(k)|^2$ is the momentum distribution; $|u(x)|^2$ is the intensity distribution, and $U_{tot}$ is the total energy of the incoming signal. For a fully coherent light source, the WDF's Fourier-transformed function $\Gamma (x) = u(x+a) u^{*} (x-a)$ is known as the mutual intensity used in wavefront sensing for turbulence analysis and adaptive detection techniques. For brevity, a list of useful optical properties of the WDF can be found elsewhere, e.g. \cite{bastiaans2009wigner}, \cite{alieva2012wigner}.

Now, let us consider a Gaussian beam, expressed in the following form \cite{galvez2006gaussian}:
\begin{equation}
    u_{G}(\rho) = \frac{A}{\text{w}} e^{-\frac{\rho^2}{\text{w}^2}}\;e^{\frac{i \tilde{\pmb{ k}}\rho^2}{2R}}
    \label{06/01_1}
\end{equation}
where $\rho>0$ is the position-vector in the beam profile $\rho = \sqrt{x^2+y^2}$, $\text{w}=\text{w}(z)$ is a beam waist, $R=R(z)$ is the radius of curvature of the beam's wavefront, and A is the normalization constant. The wave-number $\tilde{\pmb k} = \{\tilde k_x,\tilde k_y,\tilde k_z\}$ is distinctive from the Fourier transform parameter $k=\{k_x, k_y, k_z\}$ in \eqref{08_13_1}. The corresponding 1D WDF is \cite{bastiaans1997application}
\begin{equation}
W^{(G)}(\tilde x,\kappa_x) = \frac{A}{\sqrt{2\pi}} e^{-\tilde x^2 - \kappa_x^2}
    \label{06/02_1}
\end{equation}
which is plotted in of position-momentum space $\{x, k_x\}$ in Fig. \ref{fig:experiment}a. Here $\bar x = \sqrt{2} x/\text{w}$ and $\kappa_x$ is the redefined momentum:
\begin{equation}
    \kappa_x = \frac{\text{w}}{\sqrt{2}}\Big(k_x -\frac{\tilde k_x\;x}{R}\Big)
    \label{kappa_eq}
\end{equation}
This Wigner distribution is properly normalized, delivering the beam intensity distribution when integrated over the entire momentum space.

Classical Shannon information, see \cite{shannon1948mathematical}, represents the amount of structure in the corresponding system. Its analog for an optical mode, characterised by its WDF, is:
\begin{equation}
    S = -\iint_{\mathds{R}^2}d\pmb r \; d\pmb k\; W(\pmb r, \pmb k) \cdot \ln [W (\pmb r, \pmb k)]
    \label{08/11_1}
\end{equation}
where the WDF is a scalar function of position $\pmb r$ and momentum $\pmb k$ vectors. Similar definitions of information applied to characterizing optical fields, while having been suggested in the field of optics earlier, e.g. \cite{bastiaans1997application, nuttall1990two}, have not been explicitly applied to topological optical beams in the way introduced here, to the best of the authors knowledge. The main problem with this definition, eqn. \eqref{08/11_1}, is that the WDF is not a positive-semidefinite function and, hence, does not represent a proper distribution. Negativity of the WDF can be interpreted as a marker of a phase-space interference and non-classicality \cite{dragoman2005applications, kenfack2004negativity}. However, in quantum descriptions interference is associated with a local violation of Heisenberg inequality, e.g. \cite{claasen1981time}. For these reasons we seek a definition that would comply with the quantum nature of light. 

\begin{figure}
\includegraphics[scale=.48]{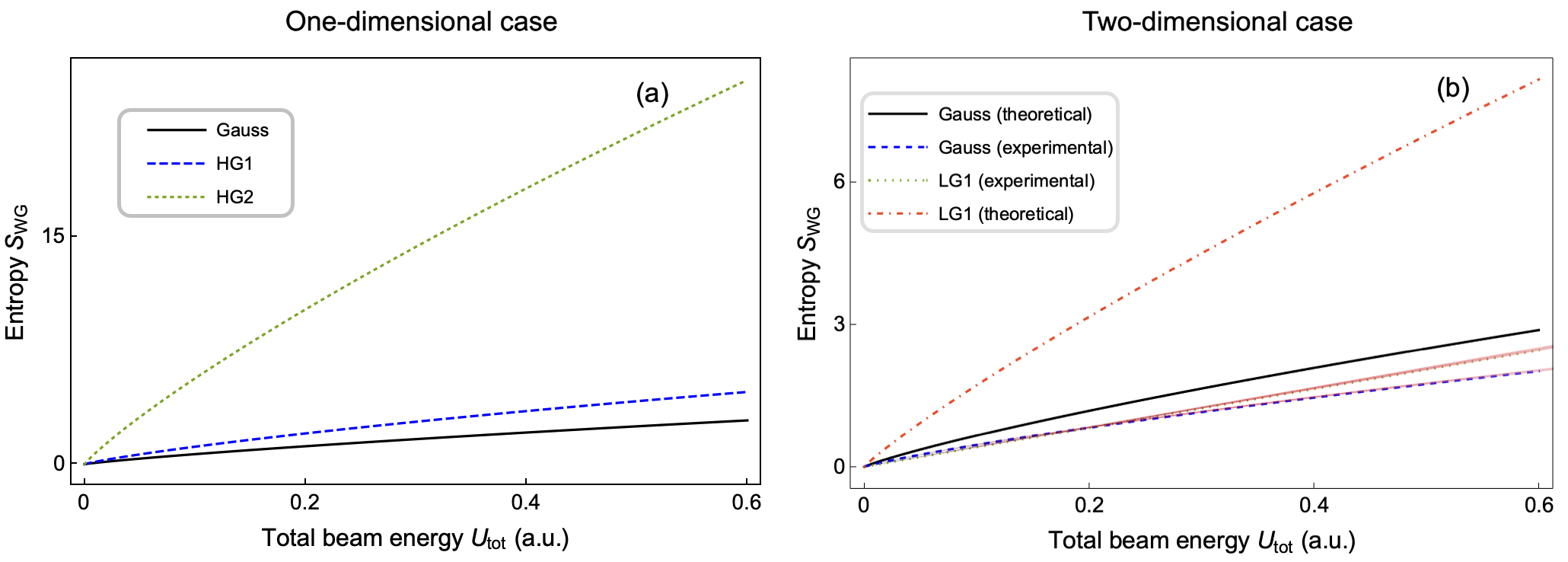}
\caption{The Shannon-Groenewold information $S_{\text{WG}}$ as a function of total energy  $U_{\text{tot}}$ \eqref{08/11_2} for one-dimensional (a) HG$_0$/Gauss mode (black), HG$_1$ (dashed-yellow) and HG$_2$ (dotted-green); and two-dimensional (b) theoretically predicted (black) and experimental (dashed-blue) Gauss mode, experimentally measured (dotted-green) and  theoretically predicted (dot-dashed-red) LG$_1$ with the corresponding error bands resulting from the errors on fit parameters, see Sec.\ref{sec: Experiment} for details. One notices the overall tendency for the amount of information to increase with the growth of the overall complexity of the corresponding optical signal. Also, the drop in the amount of information inferred from the experiment as compared to the theoretical curve is attributed to the SLM's beam conversion efficiency and expected information loss during the propagation in free space.}
\label{fig:Log_S}
\end{figure}

Here we propose to replace the regular product in the definition of information \eqref{08/11_1} with the Groenewold associative product \cite{groenewold1946principles}
\begin{equation}
\star = e^{i \hbar(\cev{\partial_x}\vvec{\partial_p} - \cev{\partial_p}\vvec{\partial_x})/2}
\end{equation}

and hereby call it Shannon-Groenewold information:
\begin{equation}
    \tilde S = -\frac{1}{(\pi \hbar)^2}\iint_{\mathds{R}^2}d\pmb r \; d\pmb k\; W(\pmb r, \pmb k) \star \ln [W (\pmb r,\pmb k)]
    \label{08/11_2}
\end{equation}
where $W(\pmb r, k)$ is the four-dimensional (4D) WDF and ($\star$) is the Groenewold's associative product \cite{curtright2001negative}. A manifestation of this definition is the Weyl quantization of a classical observable in phase-space \cite{mcdonald1987phase}. It respects the Heisenberg uncertainty principle and is normalized to the phase space volume. As an entropic observable, it measures the structure present in the optical field analogous to \cite{khinchin1953entropy}. In the context of Information Theory, it may be interpreted as a measure of the amount of \emph{classical} information that can be extracted from the laser mode if all the quantum uncertainty is removed by an appropriate experiment.

Using the Shannon-Groenewold information \eqref{08/11_2} we obtain the following expression for the information in the Gaussian beam:
\begin{flalign}
    \tilde S_{1D} \simeq -U_{tot} \Big[\ln \Big(\frac{U_{tot}}{\pi}\Big)-1\Big] \nonumber \\
    \tilde S_{2D} \simeq -U_{tot}^2 \Big[\ln \Big(\frac{U_{tot}^2}{\pi^2}\Big)-2\Big]
    \label{06/02_2}
\end{flalign}
where $U_{tot} = A\sqrt{\pi/2}$ is the total EM-energy of the beam per spatial degree of freedom. We assume that the radius of curvature of the beam wavefront is always larger than the physical dimensions of the beam spot size: $R \gg \{ x, y \}$. It is important to note that for the Gaussian mode both definitions \eqref{08/11_1} and \eqref{08/11_2} produce the same result. This is what one would expect since the WDF of a pure Gaussian sources is positive-semidefinite and can be interpreted as a classical proper distribution function \cite{dragoman2000phase}.

\section{Higher Order Modes and Shannon-Groenewold information} \label{sec: Higher order}

The HG mode is a higher-order solution of the Gaussian family of beam-like solutions of the paraxial equation. It carries a non-trivial beam geometry with singularities, stable on propagation \cite{gbur2007singular} due to being topologically protected. Hence, it is only logical to expect that the amount of information for this family of modes is higher than for the zero-order Gauss modes $\tilde S_{G} \lesssim \tilde S_{HG}$ \cite{kumar2011information}. Let us investigate this statement next. 

The general solution of the 1D paraxial wave equation in cylindrical coordinates is given by
\begin{equation}
    u_{HG}(x,z) = \sqrt{\frac{A}{\text{w}}} H_m \Big( \frac{\sqrt{2}x}{\text{w}} \Big) e^{-x^2/\text{w}^2} e^{i\tilde k_x x^2/2R}
    \label{28_10_3}
\end{equation}
The normalization is consistent with the one in eqn. \eqref{06/01_1}. The WDF for this mode is known, e.g. \cite{simon2000wigner}. In our case, to keep the normalization consistent, we obtain:
\begin{flalign}
    W^{(HG)}_{m}(x,k_x) = \frac{A\sqrt{\pi}}{(2\pi)^{3/2}} e^{-\bar{x}^2}  \int_{-\infty}^{\infty} d\bar{y} \;e^{-i\kappa_x \; \tilde y - \frac{\bar y^2}{4}}  H_m \Big(  \bar x + \frac{\bar y}{2} \Big) H_m \Big(  \bar x - \frac{\bar y}{2} \Big)
    \label{06/03_1}
\end{flalign}
As expected, for $m=0$, the WDF of HG mode is exactly equal to the Gaussian WDF, eqn.\eqref{06/02_1}, and so is the corresponding information in eqn. \eqref{06/02_2}. Then, the first two higher order WDFs have 'elegant' analytical expressions:
\begin{flalign}
W_1^{(HG)} (x, k_x) = \frac{2A}{\sqrt{2\pi}} e^{-\bar x^2-\kappa_x^2}H_1\Big(\bar x^2+\kappa_x^2-\frac{1}{2}\Big);\label{11/10/2}\\
W_2^{(HG)}(x, k_x) = \frac{4A }{\sqrt{2\pi}}e^{-\bar x^2-\kappa_x^2}H_2\Big(\bar x^2+\kappa_x^2-1\Big).
\label{11/10/1}
\end{flalign}
The WDF $W_1^{(HG)}(x, k_x)$ is plotted in Fig. \ref{fig:experiment}b. One can see explicitly the negative contributions, in 2nd-order mode in opposition to positive-definite 0th-order Gauss beam. In a similar manner one can work out higher order modes using the integral form in eqn. \eqref{06/03_1}.

Next, we explore these WDF expressions further. We find that these WDFs take negative values in the central region of the phase-space (Fig. \ref{fig:experiment}). The fundamental statement of classical information theory that "\emph{the more we know about a system's parameter space, the less is its uncertainty}" is inevitably broken in the quantum context when considering correlated (conjugate) variables, such as position $x$ and momentum $p$. If such "quantumness" is present in the corresponding PDF, it pushes the distribution into the negative domain \cite{kenfack2004negativity}, Fig. \ref{fig:experiment}b. By respecting Weyl-Wigner quantization in the definition of information \eqref{08/11_2}, we work around the WDF not being a well-defined PDF from the statistics point of view. These lead to real and positive-valued, monotonously increasing with energy information:
\begin{flalign}
\tilde S^{(HG)}_{1(1D)} \simeq -2U_{tot}\Big[ \ln \Big( \frac{2U_{tot}}{\pi} \Big)-3 \Big]; \label{S_HG_1}\\
\tilde S_{2(1D)}^{(HG)} \simeq -8U_{tot}\Big[ \ln \Big(\frac{4U_{tot}}{\pi}\Big)-5 \Big]. \label{S_HG_2}
\end{flalign}
As HG modes form a complete orthonormal set, they can be used as an expansion basis \cite{kimel1993relations}. Hence, this approach can be straightforwardly applied to OAM modes, such as Laguerre-Gauss (LG), e.g. \cite{galvez2006gaussian}. Let us express the LG mode as follows:
\begin{flalign}
u_{LG}(\rho,z) =\frac{A}{w} \Big( \frac{\sqrt{2} \rho}{w} \Big)^{|\ell|} L_{p}^{|\ell|} \Big(\frac{2\rho^2}{w^2}\Big) e^{-\rho^2/w^2} e^{ik\rho^2/2R} e^{i\ell \phi}
\label{28_10_2}
\end{flalign}
where $\ell$ is the vorticity of the twisted mode; $\phi$ is the azimuthal angle in cylindrical coordinates, where the remaining parameters follow the definitions of Gauss \eqref{06/01_1} and HG \eqref{28_10_3} modes. Supplied the results from \eqref{06/03_1}, we can express LG-modes in terms of HG-modes with a straightforward calculation; for instance, for the WDF of 2D LG mode with $p=0$ and $\ell = 1$:
\begin{flalign}
W_{1(2D)}^{(LG)} &(\bar \rho; \kappa) =\frac{A^2}{2 \pi} e^{-{\bar \rho}^2 - \kappa^2} \left(\bar \rho^2 +\kappa^2-1\right)
\label{28_10_11}
\end{flalign}
where $p$ and $\ell$ are correspondingly the order- and degree - numbers of the generalized Laguerre polynomial $L_{p}^{\ell}(\boldsymbol{\cdot})$; $\bar \rho=\sqrt{\bar x^2+\bar y^2}$. The corresponding entropy can be calculated in a similar manner:
\begin{flalign}
\tilde S_{1 (2D)}^{(LG)} \simeq -2 U_{tot}^2 \left( \ln\left( \frac{U_{tot}^2}{\pi^2} \right) -4 \right)
\label{22_11_1}
\end{flalign}
The classical assessment of the amount of order in the optical mode is clearly increasing with the increasing complexity of the beam profile, see Fig. \ref{fig:Log_S}, as one would expect from general considerations. It is important to recall that in the context of physical meaning of Shannon unconditional information, the WDF is normalised to $U_{tot}$ -- total energy. Consequently, constant $A$ is bounded from above by $\sqrt{2/\pi}$.

The similarity sign in the expressions for information (structural complexity) of considered above modes \eqref{06/02_2}, \eqref{S_HG_1}, \eqref{S_HG_2}, and \eqref{22_11_1} is due to the presence of terms containing an odd logarithmic integral of the form:

\begin{flalign}
\int_0^{\infty} dr\; r\;\ln(2r^2-1) I_0(\alpha r) \rightarrow 0
\end{flalign}
where $I_0$ is a modified Bessel function of the first kind. Numerical estimations show tendency for these terms to go to zero, however, mathematically rigorous study of their absolute convergence has not been performed.

With this theoretical framework, we are ready to perform an experiment to explore the possibility of obtaining the amount of information by measuring the structure in optical laser beams of various topology.
\begin{figure}
\includegraphics[scale=0.18]{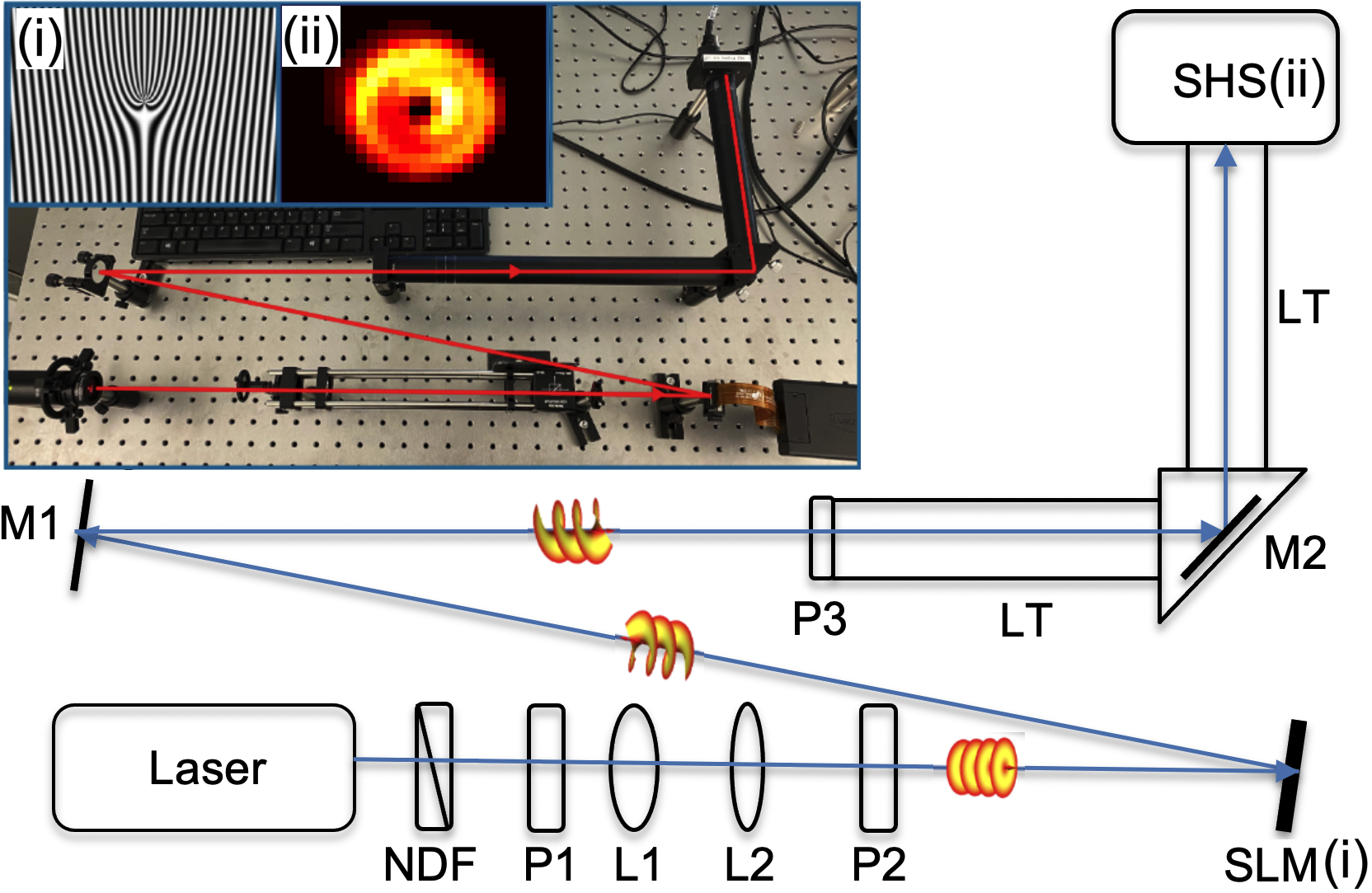}
\caption{The schematic of the experimental setup, where a spatially filtered Gauss-like 635nm, $4$mW (Thorlabs LDM635) laser beam is reflected off the spatial light modulator (SLM) onto the mirror M1, into the boxing of the Shack-Hartmann sensor (SHS), Thorlabs: WFS20-7AR. The device is optimized to operate with ordinary Gauss-like signals. The SLM can be set to a mirror regime or to generate an Orbital Angular Momentum (OAM) beam of Laguerre-Gauss-like profile. The OAM beam generation is accomplished by an SLM loaded with a computer-generated diffraction pattern with a fork dislocation (i). In that case the measured intensity distribution of an OAM beam (ii) has a typical doughnut-like structure. L1 and L2 are the lenses of the beam expander; P1,2 and 3 are the pinholes; M1 and 2 are the directing mirrors; LT are the boxing elements of the SHS.}
\label{fig:setup}
\end{figure}

\section{Experimental approach} \label{sec: Experiment}
Among the tools of adaptive optics, Shack-Hartman sensors (SHS) \cite{platt2001history} occupy a unique place as a fast, affordable, compact off-the-shelf tool for simultaneous intensity and angular distribution measurements. Advanced techniques for SHS state tomography \cite{stoklasa2014wavefront} and WDF reconstruction \cite{tian2013wigner} have been suggested alongside with conventional aberration correction techniques. Measurements of the wavefront distortions in EM beams with non-trivial topology are also of interest for both communication and sensing purposes, e.g. \cite{zhao2012aberration}.

For the aim of this work, we use the SHS to reconstruct the WDF for the purpose of discriminating between the modes, assessing the beam quality, and ultimately the amount information in a beam. While a SHS is utilized generally to facilitate beam alignment and assess distortions within an optical channel by calculating the higher order Zernike moments, here we omit the SHS's wavefront data output and only consider the raw data of the intensity distribution in the superpixel array of the SHS's camera (Fig. \ref{fig:setup}). We compare the results to the modelled intensity distribution based on the theoretical WDF calculation, whose approximation can be modelled as \cite{tian2013wigner}
\begin{flalign}
I (\pmb r) = \frac{1}{\lambda f}\sum_{\substack{\ell=-L \\ m=-M}}^{L,M} SWDF[W_b, W_a] (r'_{\ell, m}, u'_{\ell, m}) \text{rect}(r'_{\ell, m})
\label{09/09/2020_1}
\end{flalign}
The smooth WDF is defined following \cite{zhang2009wigner}:
\begin{flalign}
SWDF[W_b, W_a] (r'_{\ell, m}, k'_{\ell, m}) = \iint d^2 R \;d^2 U W_b (R, U) W_a (R-r'_{\ell, m}, U-u'_{\ell, m})
\label{11/03/01}
\end{flalign}
where the coordinate shift is defined as
\begin{flalign}
&r'_{\ell, m} = \{R_x - \ell w, R_y - m w\}\\
&u'_{\ell, m} = \Big{\{}U_x - \frac{x - \ell w}{\lambda f}, U_y -\frac{y - m w}{\lambda f}\Big{\}}
\end{flalign}
The functions $W_b$ and $W_a$ are the WDF of the incoming signal, e.g. (\ref{06/02_1}, \ref{11/10/2}, \ref{11/10/1}) or \eqref{28_10_11}, and the transmission function of a single lens aperture correspondingly. The parameters $f$ - focal length, $w$ - width of a single lens in a lenslet array are the parameters specific to the detector and defining the angles in the local wavefront of the field.

\begin{figure}
\includegraphics[scale=0.26]{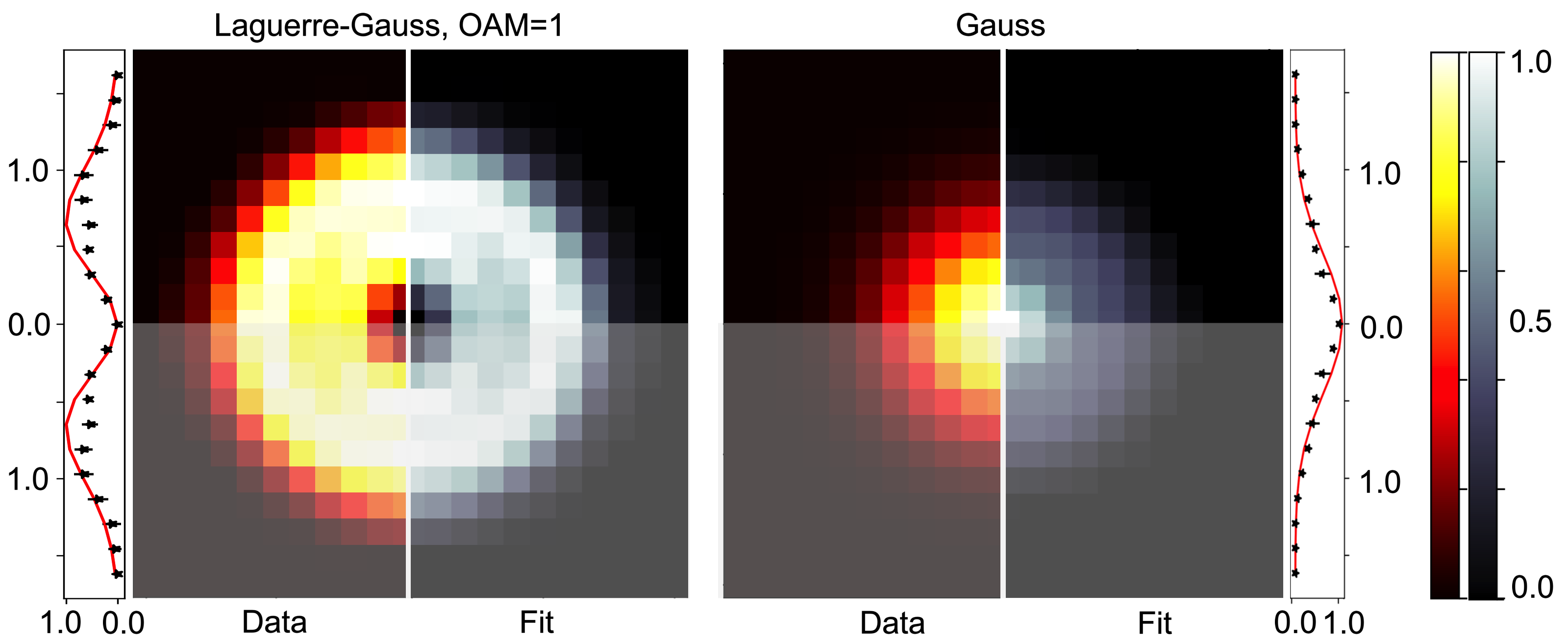}
\caption{On the way to measuring the amount of information in optical beams. The intensity distributions of the EM beams are captured by a SHS. The algorithm for simulated data-sets and the fits is based on Eqs. \eqref{09/09/2020_1} and \eqref{11/03/01}. The experimental data are averaged over 4 Laguerre-Gaussian LG$_1$ samples (left) and 10 Gaussian (right) samples with the SHS's sampling rate of 18fps (black starred scatter-plot), and over the four quadrants in the beam intensity profile. The measurement fits are shown in solid-red curves with the Gaussian model (right) using the WDF in Eq.\eqref{06/02_1}, and Laguerre-Gauss (LG) model (left) eqn. \eqref{28_10_11}. The measured data are shown as black stars with the corresponding error-bars.}
\label{fig:Fit}
\end{figure}

Using this model, we first simulated the synthetic data sets for Gauss, HG and LG modes. In the model we considered the WDF $W_b$ in the following general form:
\begin{equation}
W(\tilde x,\kappa_x) = \frac{A}{\sqrt{2\pi}} e^{-\tilde x^2 - \kappa_x^2} \;P( \tilde x^2 + k_x^2)
\label{09/29/2020_1}
\end{equation}
where P($\alpha$) is the polynomial of $\alpha$. We started by testing two cases, namely, a Gauss mode \eqref{06/02_1} and a LG of order 1 \eqref{28_10_11}. The polynomial fit in eqn. \eqref{09/29/2020_1} was taken to be
\begin{gather}
P( \tilde x^2 + k_x^2) = a + b(\tilde x^2+\kappa_x^2)
\label{09/29/2020_2}
\end{gather}
with $a$ and $b$ being the fitting parameters (Fig. \ref{fig:Fit}). One can see that when $a=const$ and $b=0$ the fit corresponds to a Gaussian profile \eqref{06/02_1} and when $a/b=2$ the model includes  LG$_1$-like distributions \eqref{28_10_11}.

\begin{figure}
\includegraphics[scale=.8]{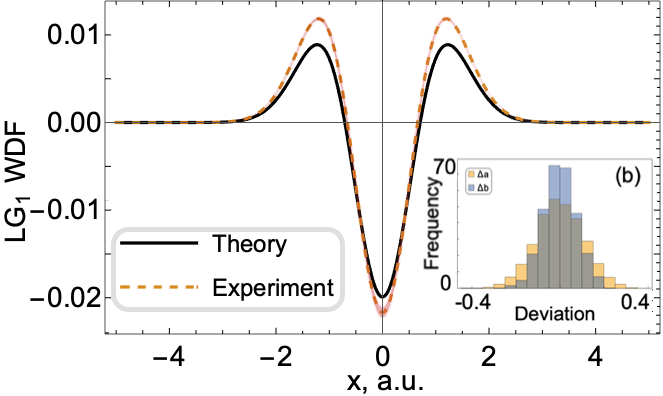}
\caption{Comparing the experiment and the theoretical prediction for the WDF of the LG mode of order $\ell=1$: (a) the WDF of the ideal LG$_1$ mode (solid black), and of the measured in an experiment laser beam (dashed yellow) with the corresponding error-band; (b) the frequency histogram of the deviation of the fitting parameters $a$ and $b$, eqn.\eqref{09/29/2020_2}, supplied to the model \eqref{09/09/2020_1}, and resulted from the fitting procedure: 1000 synthetic data sets have been generated and fitted with the model, mentioned above; for each run the deviation between the supplied and fitted parameters has been calculated.
}
\label{fig:hist}
\end{figure}

To assess the quality of the model, besides estimating the $\chi^2$ per each fit, we run a simulation with 1000 fits to synthetic data with, Eq. \eqref{09/09/2020_1}, obtaining a histogram of the deviation between the supplied fitting parameters and those from the best-fit (Fig. \ref{fig:hist}b). The resulting data show that the parameters are centered near the "true" values, showing the satisfactory quality of the fit.

The inferred fundamental amount of information carried in an experimentally measured photon beam appears to be lower than the theoretically predicted for the ideal LG$_1$ mode, see Fig. \ref{fig:Log_S}. The resulting entropy as a function of beam energy is extremely sensitive to the fit parameters $a$ and $b$ \eqref{09/29/2020_2}. 

Since we did not consider the wavefront readings of the SHS, this model does not harvest the information stored in the reciprocal domain at this point, but rather infers it from the intensity distribution. This capability is to be explored and utilized in our future research. However, even at this conceptual level, the fit already can discriminate between the two modes. Hence, it provides an experimental estimate for the WDF (Fig. \ref{fig:hist}a) supplied, however, a "good" guess about the possible wavefront shape of the incoming signal. Due to intensity-only detection, the central region of the LG beam, generated in the experiment, is left out. Hence, the setup is inherently classical, in compliance with the definition of information, used here.

The quality of the fit, alongside with the accuracy of the numerical integration algorithm, also depends on the detector modelling scheme, \eqref{09/09/2020_1} and \eqref{11/03/01}, which in this case has been fairly generic. One of the crucial assumptions that has been made is the plane-wave approximation. This approximation, generally speaking, is too brave for the case of topological beams. Another unaccounted source of discrepancies is the optical cross-talk due to SHS's architecture, that has been extensively discussed before on the level of mathematical modelling \cite{tian2013wigner}. Hence, this model, thought already fruitful, has a great potential for improvement.

These results were used to assess the information stored in a physical channel and to compare them to theoretical curves (Fig. \ref{fig:Log_S}). As for applications, the full 2D model can also provide information about the medium the beam interacted with, that can be useful in remote sensing. Based on the results of Sec. \ref{sec: Higher order}, in the course of future research, we expect topological beams to outperform the modes with planar phase structure for two main reasons: 1) greater library of non-trivial signatures in the original beam profile; 2) reported robustness and self-healing properties of vortex modes.
\section{Discussion and Summary} \label{sec: Summery}

Interestingly, while information technologies have seen outstanding progress over the last century which lead to the digital revolution and created flourishing businesses, the field of information theory has remained being relatively under-explored. We believe, that a universal technique to assess both the quality and quantity of information in a received signal, if proven, could provide a conceptually novel tool to physicists and engineers alike. The approach described in this work is by far not the first attempt, neither is it the most general. However, in this approach, classical information does not require early choice of a communication scheme (i.e. alphabet). It is rather based on a fundamental assessment of an optical system's capability to carry information, based on its overall complexity. The WDF is uniquely used here as a probability density function for Shannon information in optics. The constraints of probability theory on the definition of information and of quantum mechanics on conjugate observables are satisfied working around the properties of the WDF -- a \emph{pseudo}-probability distribution.
We foresee the relevance of this formalism in the context of recent developments for; (i) free-space information processing optics \cite{miscuglio2020massively}; (ii) integrated photonics-based information processing \cite{de2020primer} such as neural network-based accelerators \cite{peng2020dnnara} and photonic tensor cores  \cite{miscuglio2020photonic}; (iii) adaptive sensing \cite{ren2014adaptive}; and (iv) analog optical and photonic processors \cite{anderson2020roc,suninduced,miscuglio2020analog}. As the data compression coefficient is naturally bounded by Shannon information, carried by the beam \cite{preskill1998lecture}, this work indirectly points towards higher information capacity in beams with non-trivial structure, like HG, LG, Bessel-Gauss modes etc. Furthermore, due to the intrinsic connection between optical information theory and computer vision \cite{chaitin1990information, ruiz2009information}, our approach may end up being a source of new applications and tools in engineering designs of future generations of artificial intelligence.

Due to the WDF's relation to the EM-field correlation function, we foresee our approach to be extremely useful in adaptive optics. The reconstruction algorithm, when fully developed, has the potential to characterize the effects of decoherence in turbulent media, the 2D ambiguity function, time-resolved frequency distribution, alongside with commonly available corrections for aberration, astigmatism, peak-valley and rms deformation provided by the SHS measurements. The WDF formalism uniquely gives access to such characteristics as mutual intensity of stochastic wave fields, which is of high importance when describing partially coherent sources. These also have a potential to contribute in the fields of artificial intelligence through holography, optical encryption \cite{fang2020orbital}, and free-space communication \cite{amhoud2020space}. 

In perspective, as the demand on high-speed data transfer and streaming grows exponentially, ADSL and fiber-to-home technologies are less and less likely to satisfy even an average consumer's data hunger, not to mention business and government agencies calls. These, together with the recent advents in optical processing \cite{miscuglio2020massively}, micro- and nanofabrication \cite{ji2020photocurrent}, and OAM communications \cite{li2020increasing} put forward the mid-20th century's excitement around free-space communications in a new light. The new-generation free-space links will require coherent detection techniques to realise their potential to the fullest. We believe that our approach may result in a better understanding of which types of measurements and device architectures are needed to efficiently mine information from a free-space link.

\begin{acknowledgments} 
M.S. acknowledges valuable discussions and insightful comments from Nicholas Gorgone. M.S., J.Y., M.M., and V.J.S. acknowledge support from the Office of Naval Research (N00014-19-1-2595). A.A. acknowledges support from the U.S. Army Research Office Grant W911NF-19-1-0022.
\end{acknowledgments}

\section*{Appendix: literature review} 

For a physical system Shannon created a connection between the system's level of uncertainty, its associated amount of entropy, and the amount of information stored \cite{shannon1948mathematical}. While initially being heuristic, this inspired new paradigms in signal processing including concepts of energy and information, e.g. \cite{vedral2006introduction, gray2011entropy, woodward2014probability}. Shannon information theory was vastly generalised, see J. Goold \emph{et al.} \cite{goold2016role} for a review on information theory in the context of thermodynamics,statistical and quantum mechanics, and more recent work, e.g. \cite{savakar2020copy, goldberg2020extremal, mueller2020law}.

The WDF itself was originally introduced in quantum statistics \cite{wigner1997quantum} to describe behavior of a many-body quantum system in its position-momentum space. It is also one of the most widely used phase-space functions in quantum studies due to its close and straightforward relations to quantum mechanical observables. Indeed, Shannon entropy and its WDF-based definition has been used as a measure of structure in phase space in quantum mechanics \cite{dehesa2005information,laguna2010shannon,de2019shannon} with  possible implications in chemistry \cite{nascimento2018shannon} and biology \cite{hsu2017entropy}. 

In optics, the WDF was first introduced in the context of partial coherence theory and radiometry \cite{dolin1964beam,walther1968radiometry}. In this statistical picture of light fields, the WDF uniquely comprises the information stored both in real \emph{and} reciprocal domains $\{\pmb r, \pmb k\}$. This property makes it a useful tool for effects exploiting interference; in image reconstruction, e.g. \cite{liu2020role}; and in optical beam characterization, e.g. \cite{schafer2011beam, alieva2012wigner}. Relatively recently, the WDF has also been applied to communications and signal processing, e.g.  \cite{gonccalves1998pseudo,dragoman2005applications}.

Indeed, the WDF is a promising candidate to be used as a Probability Distribution Function (PDF) in information theory due to its comprehensiveness. As a measurable quantity, the WDF is a convenient tool for characterization of partially coherent sources. However, its application to structured light is non-straightforward due to negative regions that do not have a clear connection to experimentally available observables. Several approaches have been introduced to conceptually address the WDF negativity in signal processing \cite{choi1989improved}, optical coherence theory \cite{bastiaans2009wigner}, and quantum optics \cite{kenfack2004negativity}. It is important to note that negativity by itself is understood to signify non-classicality of a physical system, whether it is light or matter. The challenge, however, with the WDF breaking the non-negativity constraint only arises in attempts to build a \emph{classical} statistical description of a physical information channel.

Information theory in optics and image reconstruction originates in late 1950's -- early 1960's \cite{linfoot1955information,gabor1961iv}. Besides modern applications in communications, e.g. \cite{cheng2016channel}, it touches on the origins of computer vision and artificial intelligence (AI) \cite{chaitin1990information, ruiz2009information}. Hence, developing new fundamental approaches in optical information theory may end up being a source of new applications and tools in engineering designs of future generations of AI.
\bibliography{info_th.bib}

\end{document}